\documentclass[11pt]{letter}
\usepackage{amsmath}
\usepackage{amssymb}
\usepackage[dvips]{graphicx}
\usepackage[active]{srcltx}
\DeclareFontFamily{U}{msb}{}
\DeclareFontShape{U}{msb}{m}{n}{ <5> <6> <7> <8> <9> gen * msbm
        <10> <10.95> <12> <14.4> <17.28> <20.74> <24.88> msbm10}{}
\DeclareSymbolFont{AMSb}{U}{msb}{m}{n}
\DeclareMathSymbol{\realset}{\mathalpha}{AMSb}{"52}

\newcommand{\osd}[1]{\overset{\scriptscriptstyle #1}{)}}
\newcommand{\ose}[1]{\overset{\scriptscriptstyle #1}{(}}

\begin{document}

\begin{center}
\Large 
{\bf A Central Partition of Molecular Conformational Space. \newline
     V. The Hypergraph of $3D$ Partition Sequences.}
\end{center}

\vspace*{6mm}
\begin{center}
{\Large Jacques Gabarro-Arpa}
\end{center}

\vspace*{6mm}
\hspace*{26mm}
Ecole Normale Sup\'erieure de Cachan, LBPA,CNRS UMR 8113      \newline
\hspace*{26mm}
61, Avenue du Pr\'esident Wilson, 94235 Cachan cedex, France  \newline

\noindent \hspace*{60mm} Email: jga@gtran.org

\vspace*{4mm}
{\bf Abstract}

In a previous work a procedure was decribed for dividing the $3 \times N$-dimensional conformational space of a molecular system into a number of discrete cells, this partition allowed the building of a combinatorial structure from data sampled in molecular dynamics trajectories: the graph of cells or G, encoding the set of cells in conformational space that are visited by the system in its thermal wandering.
The information in G however, is encoded in a great number of fragments that must be aggregated.
We describe here the algorithmic procedures 1) for aggregating the information from G into an hypergraph allowing to enumerate the relevant cells from conformational space, and 2) for puttting the data in a very compact format.

\vspace*{2mm}
\hspace*{4mm} {\it Keywords}
{\bf Molecular Conformational Space, Hyperplane Arrangement, Face Lattice, Molecular Dynamics} 



\newpage
{\bf 1. Introduction}

The aim of this series of papers [1-6] is to build a set of mathematical tools for studying the energy landscape of proteins [7,8,9], and the present paper is a step further towards this goal.

The energy surface of proteins is the essential tool for understanding the physico-chemistry of basic biological processes like catalysis [9]. It is also a complex multidimensional structure that can only be built from the knowledge of the complete dynamical history of the molecule, which is currently out of reach of current molecular computing methods [9]. One reason is that the relevant molecular structures arise from combinations of small local movements within the molecular backbone that computer programs generate sequentially: the efficiency of calculations might be greatly improved if combinatorial methods could be used instead.

A big hurdle in using combinatorics for studying molecular dynamics simulations (thereafter referred as MDS) is the great accuracy needed in performing the computations (about a hundredth of angstr\"{o}m), this seems to preclude the use of discrete mathematics which form the basis of combinatorics. In fact the problem can be dealt in two steps: first one can momentarily give up numerical accuracy to study the dynamical properties of some discrete skeleton of the molecular structure, later in a second step molecular structures can be rebuilt from the skeleton [10]. The present work deals exclusively with the first part.

 The main tool developed here can be described as a \textbf{fluctuation amplifier}: the small movements of a molecular system, which are easily sampled with the current simulating tools, are encoded by means of a simple combinatorial structure, from which the set of structures corresponding to realizable combinations of these movements can be generated.

Within this approach, the $3D$-structures of protein molecules are encoded into objects called \textbf{dominance partition sequences} (DPS) [1-6], these are the generalization of a combinatorial structure known as noncrossing partition sequences [11].
They generate a linear partition of molecular conformational space\footnote[1]{For an $N$-atom molecule it is a $3 \times (N-1)$-dimensional space where each point corresponds to a $3D$ molecular conformation.}
(in what follows abridged to $CS$) into a set of connected disjoint regions called \textbf{cells}, each harboring the set of $3D$-conformations that have the same DPS.

Partitions are a useful tool for studying multi-dimensional spaces, in our case they systematically spann a much wider volume range than the set of points along a random trajectory curve generated by a MDS, they have also been used in many other contexts [7,8,11].

The aim of the preeceding papers [1-6] was to construct a graph whose nodes are the cells visited by the molecular system in its thermal wandering, two important properties of partition sequences make this construction possible :

\begin{enumerate}
 \item DPSs are \textbf{hierarchical} structures: partition sequences encoding different sets of cells can be merged into a 
       new partition sequence encoding the union set, and the process can be repeated with the new sets of cells, thus creating a
       hierarchy\footnote[2]{A structure called \textbf{partially ordered set} (\textbf{poset}). Posets are widely used tools in
       many theoretical chemistry problems [14-22].}.
       The importance of this property is that climbing the hierarchy ladder \textit{the number of cells increases exponentially
       while the sequence length increases only linearly}.
       This compact coding makes possible the construction of a graph representing huge regions of $CS$ whose size does not
       exceed the memory of a workstation computer, while keeping at the same time the essential information about the molecular
        structures.
 \item DPSs are \textbf{modular} structures: partition sequences can be decomposed into subsequences that are embedded in
       different conformational subspaces. This allows to define a \textbf{composition law}: if two partition sequences from
       two different subspaces share the same sequence for the intersection subspace, then joining both sequences gives
       a realizable sequence that corresponds to an existing set of cells [4-6].
\end{enumerate}

The first property tells us that the graph can be constructed, the second suggests how to build it: a molecular structure can be decomposed into sets of four atoms, its smallest $3D$ components, by composing the graphs of these one can build the graph of the molecule.

Atoms in MDSs are represented as pointlike structures surrounded by a force field [23,24], the convex enveloppe of a set of 4 points in $3D$-space is an irregular polytope called a \textbf{4-simplex} 
or \textbf{simplex}\footnote[3]{In what follows this denomination will be used to designate ordered sets of 4-atoms/points.}.
The conformational space of these sets is relatively small with only 13824 cells, of these only a fraction is visited by the system. With a $CS$ so small it can be plausibly assumed that the accessible cells are all visited during a MDS run.

The method for building the graph that was proposed in [2] consists in 
\begin{enumerate}
 \item Establishing a morphological classification of simplexes, where each class is defined by a set of geometrical constraints.
 \item The geometrical constraints that define a class allow to calculate the set of accessible  cells in a simplex $CS$ [4],
       thus to each class we can associate a graph where the nodes are the cells from this set with edges towards adjacent cells.
 \item On the other hand computer simulations of protein dynamics show [2,4] that in a protein structure the majority of simplexes
       evolve within a reduced number of morphologies. For each 4-atom set in the molecule the graph of its $CS$ is built by
       merging the graphs of the visited simplex morphologies.
 \item The $CS$ graph of the molecule, that was called the \textbf{graph of cells} or $\mathbf{G}$ in [4], can be built by composing
       the $CS$ graphs of the different simplexes.
\end{enumerate}

The graph of cells allows to enumerate exactly the set of kinematically accessible cells which contains
the subset of visited cells in conformational space. The region containing the dynamically interesting states can be further narrowed with the help of a generalized of dominance partition sequences (in what follows abridged to GDPS) developped in [5,6]. A structure that can be geometrically interpreted as a bouquet of cones in $CS$. Its importance lies in two facts:
\begin{enumerate}
 \item it encloses the region of $CS$ that harbours the dynamical states of the molecular system,
 \item it can be hierarchically factorised, thus the whole region can be decomposed as a product of smaller
       partitions of molecular conformational spaces, greatly reducing the algorithmic complexity involved
       in enumerating of the cells from $CS$.
\end{enumerate}

This subject is developped in the next six sections:

\begin{itemize}

 \item Section 2 contains a graphical presentation of dominance partition sequences.
 \item Section 3 discusses the factorization of generalized dominance partition sequences.
 \item Section 4 describes the procedure for factorizing the graph of cells with the GDPS.
 \item Section 5 discusses the construction of the graph, with a detailed description of a set of three
        algorithmic procedures that make the construction possible.
 \item Section 6 describes the construction of hyperlinks between sequences in the 3 dimensions of space.
\end{itemize}

There is also an appendix with the list of abbreviations used in this work.

\vspace*{4mm}
{\bf 2. The Dominance Partition Sequences}

Hidden in complex objects, like macromolecules, there are simple structures that cannot be seen because they are buried under great amounts of information. However, these structures can be made to emerge when information is selectively eliminated from the objects [12]: what remains of the $3D$-conformation is a skeleton, however all the skeleton properties are inherited by the original structure [13].

One must bear in mind that the molecular structures used in MDSs are already skeleton-structures since atoms are described as pointlike structures linked by spring-like chemical bonds and surrounded by a classical force field. A better description would be by quantum mechanical molecular orbitals, but that of course would inmediately overhelm the most powerful computers.

Before proceeding further let us point at two differences between the present approach and computer simulations: 1) the latter gives precise atomic nuclei coordinates in contrast with DPSs which give only relative atomic axis positions, 2) DPSs lend themselves beautifully to combinatorial computations, which is obviously not possible with real numbers.

Here the only information we keep from the $3D$-structure of macromolecules are the \textbf{dominance partition sequences} (DPS) [1-4], there are three such sequences: one for each cartesian coordinate $x$, $y$ and $z$. For an $N$-atom molecular system 
with atoms numbered from $1$ to $N$ the $DPS$ of a given coordinate $c$: \textit{is the sequence of atom numbers sorted in ascending order of the $c$ coordinate of their respective atoms}.

\vspace*{6mm}
\includegraphics{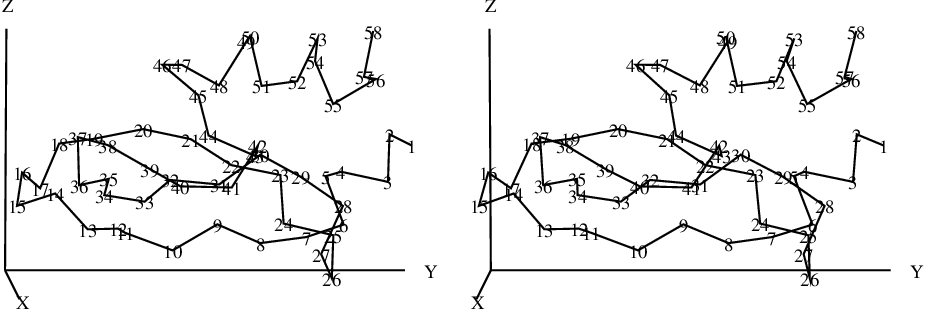}

\vspace*{4mm}
\begin{center}
\large{Figure 1} \\
\vspace*{2mm}
{\footnotesize $\alpha$-carbon skeleton stereoview of the pancreatic trypsin inhibitor [25].}
\end{center}

A simple example of $DPS$ can be extracted from Fig. 1, where an $\alpha$-carbon skeleton $3D$-conformation from the pancreatic trypsin inhibitor (PTI) [25] is shown. The associated $(x,y,z)$-dominance partition sequences of the protein C$_{\alpha}$-chain conformation are

$\{\{(58)(49)(29)(48)(57)(27)(28)(31)(30)(52) \hspace*{2mm}  (32)(47)(53)(50)(19)(26)(21)(56)(51)(24)  \newline \hspace*{4mm}
     (33)(20)(23)(55)(46)(25)(22)(34)(54)(18) \hspace*{3.5mm} (1)(45)(17) \hspace*{1.9mm}
                                                                          (5) \hspace*{1.9mm}
                                                                              (6)(35)(44) \hspace*{1.9mm}
                                                                                          (2) \hspace*{1.9mm}
                                                                                              (8)(43)  \newline \hspace*{4mm}
     (16) \hspace*{1.9mm}
          (9)(11) \hspace*{1.9mm}
                  (7) \hspace*{1.9mm}
                      (3)(10)(36) \hspace*{1.9mm}
                                  (4)(37)(42) \hspace*{2mm}  (15)(12)(41)(40)(14)(38)(13)(39)\}_x,     \newline \hspace*{2mm}
   \{(15)(16)(17)(14)(18)(37)(36)(13)(19)(38) \hspace*{2mm}  (34)(35)(12)(11)(39)(20)(33)(46)(10)(40)  \newline \hspace*{4mm}
     (32)(47)(21)(45)(44) \hspace*{1.9mm}
                          (9)(48)(31)(41)(22) \hspace*{2mm}  (49)(50)(43)(42) \hspace*{1.9mm}
                                                                              (8)(51)(30)(23)(24)(52)  \newline \hspace*{5.9mm} 
      (7)(29)(54)(53) \hspace*{1.9mm}
                      (5)(27)(55)(26)(25) \hspace*{1.9mm}
                                          (4) \hspace*{3.6mm} (6)(28)(57)(56)(58) \hspace*{1.9mm}
                                                                                  (3) \hspace*{1.9mm}
                                                                                      (2) \hspace*{1.9mm}
                                                                                          (1)\}_y,     \newline \hspace*{2mm}
   \{(26)(27)(10) \hspace*{1.9mm}
                  (8)(25) \hspace*{1.9mm}
                          (7)(24)(11) \hspace*{1.9mm}
                                      (6) \hspace*{1.9mm}
                                          (9) \hspace*{2mm}  (12)(28)(13)(33)(34)(15)(31)(32)(29)(17)  \newline \hspace*{4mm}
     (14)(23)(36)(41)(35) \hspace*{1.9mm}
                          (3)(40) \hspace*{1.9mm}
                                  (5)(22)(16) \hspace*{2mm}  (30) \hspace*{1.9mm}
                                                                  (4)(39)(43) \hspace*{1.9mm}
                                                                              (1)(18)(21)(19)(42)(20)  \newline \hspace*{4mm}
     (44)(38) \hspace*{1.9mm}
              (2)(37)(55)(48)(45)(51)(52)(57) \hspace*{2mm}  (56)(47)(46)(54)(49)(53)(58)(50)\}_z\}$  \hspace*{16mm} (1)

This means, as can be seen from Fig. 1, that the following relations hold for the $x$, $y$ and $z$ coordinates

\hspace*{18mm}
  $x_{58}$ \hspace*{0.1mm}
$< x_{49}$ \hspace*{0.1mm}
$< x_{29}$ \hspace*{0.1mm}
$< x_{48}$ \hspace*{0.1mm}
$< x_{57} <$ ...
$< x_{40}$ \hspace*{0.1mm}
$< x_{14}$ \hspace*{0.1mm}
$< x_{38}$ \hspace*{0.1mm}
$< x_{13}$ \hspace*{0.1mm}
$< x_{39}$ \newline
\hspace*{18mm}
  $y_{15}$ \hspace*{0.3mm}
$< y_{16}$ \hspace*{0.5mm}
$< y_{17}$ \hspace*{0.5mm}
$< y_{14}$ \hspace*{0.5mm}
$< y_{18} <$ ...
$< y_{56}$ \hspace*{0.3mm}
$< y_{58}$ \hspace*{0.9mm}
$< y_{3}$  \hspace*{2.0mm}
$< y_{2}$  \hspace*{1.8mm}
$< y_{1}$  \hspace*{11.5mm} (2) \newline
\hspace*{18mm}
$z_{26}$   \hspace*{0.7mm}
$< z_{27}$ \hspace*{0.7mm}
$< z_{10}$ \hspace*{0.7mm}
$< z_{8}$  \hspace*{2.0mm}
$< z_{25} <$ ...
$< z_{54}$ \hspace*{0.2mm}
$< z_{49}$ \hspace*{0.7mm}
$< z_{53}$ \hspace*{0.7mm}
$< z_{58}$ \hspace*{0.7mm}
$< z_{50}$ \hspace*{16mm}

DPSs like (1) generate an equivalence relation: two $3D$-conformations are equivalent if they have the same dominance partition sequence. Furthermore, for a $N$-atom molecular system DPSs generate a partition of the $(3 \times N - 3)$-dimensional molecular conformational space into cells whose points ($3D$-conformations) have all the same DPS. This partition is known to combinatorialists as a Coxeter reflection hyperplane arrangement and for an $N$-atom molecule is designated as 
$\mathcal{A}^{N-1} \times \mathcal{A}^{N-1} \times \mathcal{A}^{N-1}$ [26,27].

For clarity purposes we have only taken into consideration the $\alpha$-carbon atoms from the protein. Notice that this does not matter much, since the procedures used throughout this work are strictly modular and the results obtained for parts or components are also valid for the whole molecule.

As it has been extensively discussed in [11,27] these sequences have interesting combinatorial properties. Suppose we have two molecular conformations that for some coordinate axis the atoms, say 3 and 10, get past each other, obviusly these two conformations will have DPSs (encoding $(N-1)$-dimensional cells in $CS$ [4]) that differ in only two consecutive positions: $\{... (3)(10) ...\}_c$ and $\{... (10)(3) ...\}_c$ respectively.

These can be aggregated in a new sequence $\{... (3 , 10) ...\}_x$ representing the permutations of $3$ and $10$ and  encoding an $(N-2)$-dimensional cell in $CS$ [4,6].

 More generally a DPS with a sequence of $n$ atom numbers enclosed in parenthesis :

\hspace*{57mm}  $\{... (i_1 , i_2 , ... , i_{n-1} , i_n) ...\}_c$ \hspace*{51mm} (3)

 represents the set of $n!$ DPSs corresponding to the permutations of the indices
 $i_1, i_2, ... i_{n-1}, i_n $ and encodes an $(N-n)$-dimensional cell in $CS$. We call it a \textbf{permutation sequence} (abridged to PS).

\vspace*{4mm}
{\bf 3. Decomposing the Graph of Cells with Generalized Dominance Partition Sequences.}

To analyse molecular simulations with this procedure DPSs codes need to be further generalized in order to handle more complex situations. The sequence below is a valid example of the generalization we try to achieve

\hspace*{52mm}
$\{\ose{ 1} 49 \
            48 \
   \ose{ 2} 29 \
            27 \
            28 \osd{ 1} \
            30 \
            31 \
            52 \osd{ 2} \}_c$  \hspace*{46mm} (4)

where (4) encloses $\{(49 \ 48 \ 29 \ 27 \ 28)(30 \ 31 \ 52)\}_c$ and $\{(49 \ 48)(27 \ 28 \ 30 \ 31 \ 52)\}_c$ as subsequences. This means, for instance, that (4) encodes a set of cells from $CS$ where the $c$-coordinates of atom pairs $27$ and  $48$, $27$ and $31$ can be permuted, but not $48$ and $31$.

It was shown in [6] that the DPSs from the conformations generated in a molecular dynamics trajectory of the PTI protein [28] like (1), are all subsequences of the generalized DPS ($\mathbf{GDPS}$)

\vskip 3mm
$\{\{\ose{ 1} 49 \
     \ose{ 2} 48 \
     \ose{ 3} 29 \
     \ose{ 4} 27 \
              28 \
              30 \
     \ose{ 5} 31 \osd{ 1} \
              52 \osd{ 2} \
              47 \
     \ose{ 6} 32 \
     \ose{ 7} 53 \osd{ 3} \
              50 \osd{ 4} \
     \ose{ 8} 26 \osd{ 5} \
              51 \osd{ 6} \
              21 \
              23 \
              24 \
     \ose{ 9} 19 \
     \ose{10} 20 \
     \ose{11} 25 \
              33 \osd{ 7} \
              46 \
              55 \
     \ose{12} 54 \osd{ 8} \
              22 \osd{ 9} \ \newline
\hspace*{4mm} 18 \
              34 \osd{10} \
              45 \osd{11} \
     \ose{13} 17 \osd{12} \
     \ose{14}  5 \
              44 \
     \ose{15}  8 \
     \ose{16}  6 \osd{13} \
              35 \
              43 \
     \ose{17}  9 \osd{14} \
              16 \
     \ose{18} 11 \osd{15} \
               7 \osd{16} \
              36 \
     \ose{19}  3 \
               4 \
     \ose{20} 10 \osd{17} \
              42 \
     \ose{21} 37 \osd{18} \
     \ose{22} 15 \osd{19} \
     \ose{23} 12 \osd{20} \ \newline
\hspace*{4mm}
     \ose{24} 41 \osd{21} \
              40 \osd{22} \
              38 \
     \ose{25} 14 \osd{23} \
              13 \osd{24} \
              39 \osd{25}\}_{x}$ , \newline
\hspace*{1mm}
  $\{\ose{ 1} 15 \
              16 \
     \ose{ 2} 17 \osd{ 1} \
     \ose{ 3} 14 \osd{ 2} \
     \ose{ 4} 18 \osd{ 3} \
     \ose{ 5} 36 \
     \ose{ 6} 13 \osd{ 4} \
              37 \osd{ 5} \
     \ose{ 7} 19 \
     \ose{ 8} 34 \osd{ 6} \
              12 \
              35 \
              38 \osd{ 7} \
     \ose{ 9} 11 \osd{ 8} \
              20 \
              33 \
     \ose{10} 39 \osd{ 9} \
     \ose{11} 46 \
     \ose{12} 10 \osd{10} \
              32 \
              40 \
              47 \osd{11} \ \newline
\hspace*{4mm}
     \ose{13} 21 \osd{12} \
     \ose{14} 45 \osd{13} \
              44 \
     \ose{15} 31 \
     \ose{16}  9 \
              48 \osd{14} \
              41 \osd{15} \
     \ose{17} 22 \osd{16} \
     \ose{18} 42 \
              49 \
              50 \osd{17} \
               8 \
              30 \
              43 \
              51 \osd{18} \
     \ose{19} 23 \
     \ose{20} 24 \osd{19} \
     \ose{21}  7 \
              52 \
     \ose{22} 29 \
     \ose{23}  4 \
              53 \
              54 \osd{20} \ \newline
\hspace*{4mm}
     \ose{24} 26 \
              27 \osd{21} \
               5 \osd{22} \
               6 \
              25 \
              28 \
              55 \osd{23} \
               3 \osd{24}\}_{y}$ , \newline
\hspace*{1mm}
  $\{\ose{ 1} 26 \
     \ose{ 2} 27 \
     \ose{ 3}  8 \
              10 \
     \ose{ 4}  7 \
     \ose{ 5} 25 \
     \ose{ 6} 11 \
     \ose{ 7} 13 \osd{ 1} \
               9 \osd{ 2} \
               6 \
              24 \
     \ose{ 8} 12 \
     \ose{ 9} 28 \osd{ 3} \
              33 \osd{ 4} \
              31 \osd{ 5} \
     \ose{10} 34 \
     \ose{11} 15 \osd{ 6} \
              32 \
     \ose{12} 29 \
     \ose{13} 17 \osd{ 7} \
     \ose{14} 14 \osd{ 8} \
               5 \
              23 \
     \ose{15}  4 \ \newline
\hspace*{4mm}
              22 \
              35 \
              36 \
              40 \
              41 \
     \ose{16}  3 \osd{ 9} \
              30 \osd{10} \
              39 \osd{11} \
              16 \
              21 \
              43 \osd{12} \
     \ose{17} 38 \osd{13} \
              18 \
              19 \osd{14} \
              20 \osd{15} \
              37 \
              42 \
              44 \osd{16} \
     \ose{18} 55 \osd{17} \
     \ose{19} 45 \
              48 \
     \ose{20} 52 \osd{18} \
              51 \osd{19} \ \newline
\hspace*{4mm}
     \ose{21} 46 \
              47 \osd{20} \
              54 \
     \ose{22} 49 \
              53 \osd{21} \
              50 \osd{22} \}_{z}\}$ \hspace*{98mm} (5)

a graphical, more intuitive form of (5) can be seen in Fig. 2, where the permutation sequences are enclosed within squares to emphasize the intersections among them.

The $\mathbf{GDPS}$ is a $CS$ cone boundary for the dynamical states of a molecular system [27], it greatly
simplifies the algorithmic problem addressed in this work consisting in \textit{recovering the full molecular DPSs from the 4-projections contained in the $\mathbf{G}$ graph}.

Since the permutation sequences in (5) are mere boundaries they should be distinguished from the ones arising in DPSs that correspond to real molecular conformations, hence we will call them \textbf{generalized permutation sequences}, abridged to GPS.

In order to perform calculations (5) has to be transfomed into the graphs in Fig. 3, where the nodes are the  $X$, $Y$ and $Z$ GPSs from Fig. 2, and links between two nodes arise if:
\begin{enumerate}
 \item their respective sequences have a null intersection,
 \item the succesion order of the nodes along any path is the same as in Fig. 2,
 \item no non-intersecting GPS from (5) can be intercalated in between.
\end{enumerate}

The paths in the graph from Fig. 3 are the sets of maximal non-intersecting sequences from (5), in practice this  amounts to dividing the cone (5) in $CS$ into a set of smaller cones represented by the graphical paths from Fig. 3. Furthermore, each sub-cone can be factorised into a set of lower-dimensional permutation sequence cones: the ordering of GPSs within a path corresponds to a sequence ordering, as no permutations of atom positions can occur outside the GPSs.

Otherwise stated: we need only compute the DPSs for the PS cones, the molecular DPSs are ob-

\newpage
\includegraphics{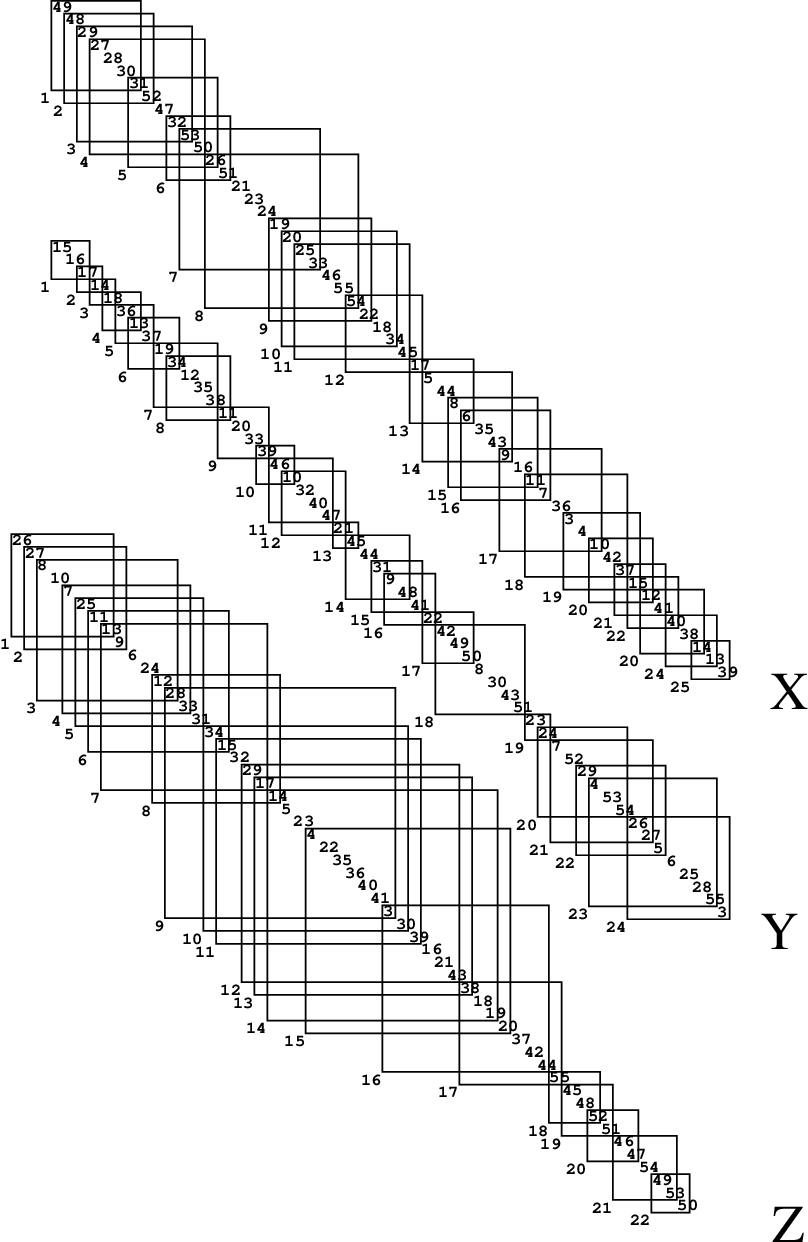}

\vspace*{4mm}
\begin{center}
\large{Figure 2} \\
\vspace*{2mm}
\footnotesize{The generalized partition function (13) in graphical form. With the permutation sequences enclosed in squares}.
\end{center}

\newpage
\includegraphics{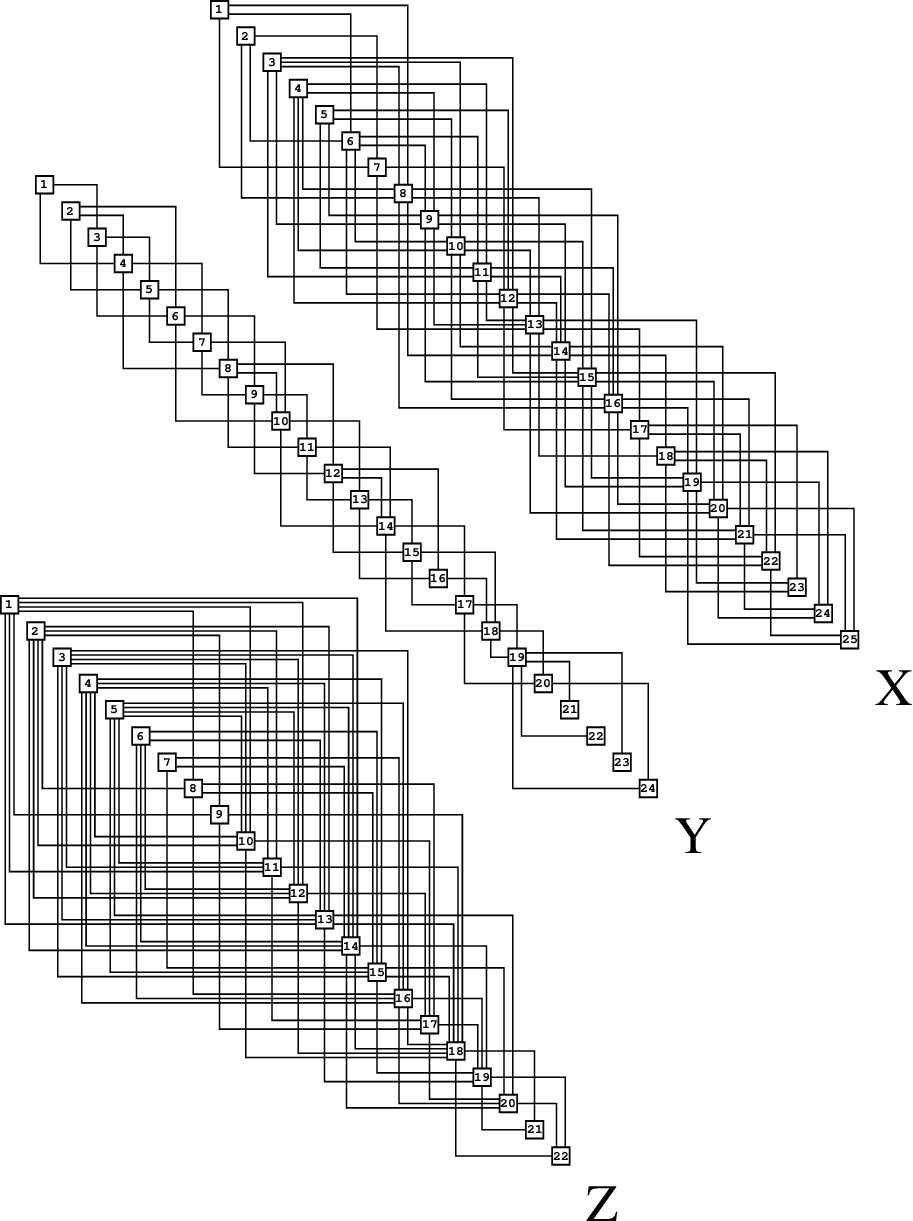}

\begin{center}
\large{Figure 3} \\
\vspace*{2mm}
\footnotesize {GDPS skeleton graph: the nodes are the permutation sequences from (5) with links towards
               adjacent non-intersecting sequences}.
\end{center}

tained by joining  the smaller ones along a path. Thus, the recovery from 4-projections can be done by restricting $\mathbf{G}$ to the permutation sequences from (5), which gives smaller and more manageable subgraphs and therefore reduces algorithmic complexity.

In this way, except for hyperlinks among sequences that will be discussed below, it suffices to calculate the DPSs for each interval 

\vspace*{4mm}
{\bf 4. Building dominance partition sequences from $\mathbf{G}$}

For each GPS-cone from Fig. 3 we restrict $\mathbf{G}$ to the 4-simplexes whith vertex numbers from the cone sequence in order to extract the set permutation sequences from the DPS-projections on the simplexes. In the example from table I we have the complete PS set for the $\mathbf{GDPS}$ node
GPS$_{x,1} = (27,28,29,30,31,48,49)$\footnote{The individual nodes from (5) and Figs. 2 and 3 are designated by two consecutive subindices: $x$, $y$ or $z$ and the node number.}

\hspace*{30mm} $(27) \ (28) \ (29) \ (30) \ (31) \ (48) \ (49) \ (27,28)$       \newline
\hspace*{30mm} $(27,29) \ (27,30) \ (27,31) \ (27,48) \ (27,49)$                \newline
\hspace*{30mm} $(28,29) \ (28,30) \ (28,31) \ (28,48) \ (28,49)$                \newline
\hspace*{30mm} $(29,30) \ (29,31) \ (29,48) \ (29,49) \ (30,31)$                \newline
\hspace*{30mm} $(30,48) \ (30,49) \ (31,48) \ (31,49) \ (48,49)$                \newline
\hspace*{30mm} $(27,28,31) \ (27,28,48) \ (27,29,30) \ (27,29,48) \ (27,29,49)$ \newline
\hspace*{30mm} $(27,30,31) \ (27,48,49) \ (28,29,30) \ (28,29,48) \ (28,30,31)$ \newline
\hspace*{30mm} $(28,30,49) \ (28,31,49) \ (29,31,48) \ (29,31,49) \ (29,48,49)$ \newline
\hspace*{30mm} $(30,31,48) \ (30,48,49) \ (31,48,49) \ (29,31,48,49)$           \newline
\hspace*{30mm} $(27,28,29,48,49) \ (27,28,30,31,49) \ (29,30,31,48,49)$

\begin{center}
\large{Table I}
\footnotesize{Set of $x$-permutation sequences obtained by restricting $\mathbf{G}$ to GPS$_{x,1}$}.
\end{center}

The example from table II will help to understand how these PSs are extracted from the 4-DPS projections in simplexes.

$\hspace*{13mm} \{27,29,48,49\} : \{\{(29,48,49)(27)\}_x,     \hspace*{1mm}
                                    \{(48)(49)(27)(29)\}_y,
                                    \{(27)(29)(48)(49)\}_z\}$ \newline
$\hspace*{13mm} \{27,28,29,48\} : \{\{(29,48)(27,28)\}_x,     \hspace*{1mm}
                                    \{(48)(27)(29)(28)\}_y,
                                    \{(27)(28)(29)(48)\}_z\}$ \newline
$\hspace*{13mm} \{27,28,29,30\} : \{\{(29)(27,28)(30)\}_x,
                                    \{(30)(27)(29)(28)\}_y,
                                    \{(27)(28)(29)(30)\}_z\}$ \newline
$\hspace*{40.4mm}                 \{\{(29)(30)(27,28)\}_x,
                                    \{(30)(27)(29)(28)\}_y,
                                    \{(27)(28)(29)(30)\}_z\}$ \newline
$\hspace*{40.4mm}                 \{\{(30)(29)(27,28)\}_x,
                                    \{(30)(27)(29)(28)\}_y,
                                    \{(27)(28)(29)(30)\}_z\}$ \newline
$\hspace*{13mm} \{27,28,30,31\} : \{\{(31)(27,28)(30)\}_x,
                                    \{(31)(30)(27)(28)\}_y,
                                    \{(27)(28,31)(30)\}_z\}$  \newline
$\hspace*{40.4mm}                 \{\{(31)(30)(27,28)\}_x,
                                    \{(31)(30)(27)(28)\}_y,
                                    \{(27)(28,31)(30)\}_z\}$  \newline
$\hspace*{40.4mm}                 \{\{(30)(31)(27,28)\}_x,
                                    \{(31)(30)(27)(28)\}_y,
                                    \{(27)(28,31)(30)\}_z\}$  \newline
$\hspace*{13mm} \{28,30,31,49\} : \{\{(31,49)(28)(30)\}_x,
                                    \{(31)(49)(30)(28)\}_y,
                                    \{(28,31)(30)(49)\}_z\}$  \newline
$\hspace*{40.4mm}                 \{\{(31,49)(30)(28)\}_x,
                                    \{(31)(49)(30)(28)\}_y,
                                    \{(28,31)(30)(49)\}_z\}$  \newline
$\hspace*{40.4mm}                 \{\{(30)(31,49)(28)\}_x,
                                    \{(31)(49)(30)(28)\}_y,
                                    \{(28,31)(30)(49)\}_z\}$  \newline
$\hspace*{13mm} \{30,31,48,49\} : \{\{(31,48,49)(30)\}_x,     \hspace*{1.1mm}
                                    \{(31,48)(49)(30)\}_y,    \hspace*{1.3mm}
                                    \{(31)(30)(48)(49)\}_z\}$ \newline
$\hspace*{40.4mm}                 \{\{(30)(31,48,49)\}_x,     \hspace*{1.3mm}
                                    \{(31,48)(49)(30)\}_y,    \hspace*{1.3mm}
                                    \{(31)(30)(48)(49)\}_z\}$ \newline
$\hspace*{13mm} \{29,31,48,49\} : \{\{(29,31,48,49)\}_x,      \hspace*{2.4mm}
                                    \{(31,48)(49)(29)\}_y,    \hspace*{1.3mm}
                                    \{(31)(29)(48)(49)\}_z\}$

\begin{center}
\large{Table II} {\footnotesize Set of $x$-projections of the permutation sequence $(29,31,48,49)$}.
\end{center}

The first column list a set of seven consecutive adjacent simplexes: each simplex from the list shares a face (i.e. 3 vertex numbers) with the preceeding and the following ones, and the same for the first and the last; also every pair of indices from GPS$_{x,1}$ can be found in at least one simplex.

The remaining columns are DPS subsets from the $CS$ of every simplex in the sequence. Before proceeding further we need the following definition

\textbf{Definition 1} : two 4-DPSs from two adjacent simplices are said \textbf{compatible} if they have the same projection on the common $3$-simplex face.

The structure of 4-DPSs in table II is such that each DPS is compatible with at least one other DPS on every adjacent simplex. Since the table has been built to include every dominance relation between nodes the projections can be composed to give the DPSs:

$\{\{(29,31,48,49)(27,28)(30)\}_x,
   \{(31,48)(49)(30)(27)(29)(28)\}_y,
   \{(27)(28,31)(29)(30)(48)(49)\}_z\}$ \newline
$\{\{(29,31,48,49)(30)(27,28)\}_x,
   \{(31,48)(49)(30)(27)(29)(28)\}_y,
   \{(27)(28,31)(29)(30)(48)(49)\}_z\}$ \newline
$\{\{(30)(29,31,48,49)(27,28)\}_x,
   \{(31,48)(49)(30)(27)(29)(28)\}_y,
   \{(27)(28,31)(29)(30)(48)(49)\}_z\}$

\begin{center}
\large{Table III} \footnotesize{DPSs resulting from composing the 4-DPSs from table II}.
\end{center}

whose $x$ components correspond, as we shall see below, to the three allowed permutations of the DPS$_x$ $(30)(27,28)(29,31,48,49)$ for the GPS$_{x,1}$. Table III gives a hint as to the enormous compaction power of permutation sequences: for the 3 partition sequences from this table the $x$ components alone encode 144 simple DPSs.

\vspace*{4mm}
{\bf 5. Building The Dominance Partition Sequences Set for One Coordinate}

Repeating the procedure described in the preceeding paragraph for the whole set 4-DPS projections we obtain the complete set of partitions sequences in GPS$_{x,1}$. These are shown in table IV up to a maximum of 5 PSs for sequence (in order to avoid an unnecessary long table).

Table IV presents a further compatification of the code: for each DPS permutation sequences are ordered
\begin{enumerate}
 \item according to their length starting on the left with the shortest one,
 \item inside a PS elements are ordered by their numerical values,
 \item equal sequences are sorted by comparing their elements in a depth first manner.
\end{enumerate}

An hexadecimal code on the right indicates the occuring sequences from the set of ordered permutations.

The code '$\mathtt{4c}$' from the third sequence in the left column of table IV: $(30)(27,28)(29,31,48,49)$, for instance, can be translated into the binary code '01001100': this means that only the permutations 2, 5 and 6 from the ordered set are found to occur, these are precisely the DPS$_x$s from table III.

The molecular DPS$_c$ for a given coordinate ($c = x \text{ , } y \text { or } z$) are calculated for each GPS for every path in the graphs from Fig. 3, partition sequences in different GDPs by construction are totally independent. Thus: \textit{the set of molecular} DPS\textit{s is the sequentially ordered product of the} DPS \textit{sets of each} GDP \textit{along a graphical path}.

\newpage
The following lemma is the formalization the previous statement.

\begin{itemize}
 \item let $\mathcal{P}_c$ be a coordinate $c$ path,
 \item let $\text{GPS}_{c,n} \in \mathcal{P}_c \ \text{for} \ 1 \leq n \leq |\mathcal{P}_c| $ be its nodes,
 \item let $\{\text{DPS}_{c,n}\}$ be the set of DPSs for node $\text{GPS}_{c,n}$,
 \item let $\{\text{DPS}\}_{\mathcal{P}_c}$ be the set of molecular DPSs of $\mathcal{P}_c$, then
\end{itemize}

\textbf{Lemma 1} $\{\text{DPS}\}_{\mathcal{P}_c} \ = \underset{1 \leq n \leq |\mathcal{P}_c|} \prod {\{\text{DPS}\}_{\mathcal{P}_c}}$

\vspace*{28mm}
\hspace*{15mm} $(27,28) \ (29,30,31,48,49) \ \ \ \ \ \ \ \ \mathtt{40}  \ \ \ \ \ \ \ \ \ \
                (30,31) \ (27,28,29,48,49) \ \ \ \ \ \ \ \ \mathtt{40}$ \newline
\hspace*{15mm} $(28) \ (27,29,30) \ (31,48,49) \ \ \ \ \ \ \mathtt{34}  \ \ \ \ \ \ \ \ \ \
                (30) \ (31) \ (27,28,29,48,49) \ \ \ \ \ \ \mathtt{0c}$ \newline
\hspace*{15mm} $(30) \ (27,28) \ (29,31,48,49) \ \ \ \ \ \ \mathtt{4c}  \ \ \ \ \ \ \ \ \ \
                (30) \ (27,28,31) \ (29,48,49) \ \ \ \ \ \ \mathtt{4c}$ \newline
\hspace*{15mm} $(30) \ (27,28,48) \ (29,31,49) \ \ \ \ \ \ \mathtt{4c}  \ \ \ \ \ \ \ \ \ \
                (31) \ (27,48,49) \ (28,29,30) \ \ \ \ \ \ \mathtt{fc}$ \newline
\hspace*{15mm} $(48) \ (27,29,30) \ (28,31,49) \ \ \ \ \ \ \mathtt{b0}  \ \ \ \ \ \ \ \ \ \
                (49) \ (27,30,31) \ (28,29,48) \ \ \ \ \ \ \mathtt{4c}$ \newline
\hspace*{15mm} $(27,28) \ (30,31) \ (29,48,49) \ \ \ \ \ \ \mathtt{0c}  \ \ \ \ \ \ \ \ \ \
                (27,28) \ (30,49) \ (29,31,48) \ \ \ \ \ \ \mathtt{1c}$ \newline
\hspace*{15mm} $(27,29) \ (28,30) \ (31,48,49) \ \ \ \ \ \ \mathtt{c8}  \ \ \ \ \ \ \ \ \ \
                (27,29) \ (31,48) \ (28,30,49) \ \ \ \ \ \ \mathtt{e0}$ \newline
\hspace*{15mm} $(27,30) \ (28,29) \ (31,48,49) \ \ \ \ \ \ \mathtt{34}  \ \ \ \ \ \ \ \ \ \
                (27,30) \ (31,49) \ (28,29,48) \ \ \ \ \ \ \mathtt{1c}$ \newline
\hspace*{15mm} $(27,31) \ (48,49) \ (28,29,30) \ \ \ \ \ \ \mathtt{fc}  \ \ \ \ \ \ \ \ \ \
                (27,48) \ (31,49) \ (28,29,30) \ \ \ \ \ \ \mathtt{fc}$ \newline
\hspace*{15mm} $(27,49) \ (31,48) \ (28,29,30) \ \ \ \ \ \ \mathtt{fc}  \ \ \ \ \ \ \ \ \ \
                (28,29) \ (48,49) \ (27,30,31) \ \ \ \ \ \ \mathtt{e0}$ \newline
\hspace*{15mm} $(28,30) \ (31,48) \ (27,29,49) \ \ \ \ \ \ \mathtt{1c}  \ \ \ \ \ \ \ \ \ \
                (28,31) \ (48,49) \ (27,29,30) \ \ \ \ \ \ \mathtt{1c}$ \newline
\hspace*{15mm} $(28,48) \ (31,49) \ (27,29,30) \ \ \ \ \ \ \mathtt{1c}  \ \ \ \ \ \ \ \ \ \
                (28,49) \ (31,48) \ (27,29,30) \ \ \ \ \ \ \mathtt{1c}$ \newline
\hspace*{15mm} $(29,30) \ (31,49) \ (27,28,48) \ \ \ \ \ \ \mathtt{e0}  \ \ \ \ \ \ \ \ \ \
                (29,31) \ (30,49) \ (27,28,48) \ \ \ \ \ \ \mathtt{e0}$ \newline
\hspace*{15mm} $(29,48) \ (30,49) \ (27,28,31) \ \ \ \ \ \ \mathtt{e0}  \ \ \ \ \ \ \ \ \ \
                (29,49) \ (30,31) \ (27,28,48) \ \ \ \ \ \ \mathtt{e0}$ \newline
\hspace*{15mm} $(28) \ (31) \ (48,49) \ (27,29,30) \ \ \ \ \mathtt{0071ff}  \ \ \ \
                (28) \ (49) \ (31,48) \ (27,29,30) \ \ \ \ \mathtt{0071ff}$ \newline
\hspace*{15mm} $(29) \ (31) \ (30,49) \ (27,28,48) \ \ \ \ \mathtt{ff8e00}  \ \ \ \
                (29) \ (49) \ (30,31) \ (27,28,48) \ \ \ \ \mathtt{ff8e00}$ \newline
\hspace*{15mm} $(30) \ (31) \ (27,28) \ (29,48,49) \ \ \ \ \mathtt{0c003f}  \ \ \ \
                (30) \ (31) \ (29,49) \ (27,28,48) \ \ \ \ \mathtt{b2cfc0}$ \newline
\hspace*{15mm} $(30) \ (49) \ (27,28) \ (29,31,48) \ \ \ \ \mathtt{4d303f}  \ \ \ \
                (30) \ (49) \ (29,31) \ (27,28,48) \ \ \ \ \mathtt{b2cfc0}$ \newline
\hspace*{15mm} $(30) \ (49) \ (29,48) \ (27,28,31) \ \ \ \ \mathtt{b2cfc0}  \ \ \ \
                (31) \ (48) \ (27,29) \ (28,30,49) \ \ \ \ \mathtt{b2cfc0}$ \newline
\hspace*{15mm} $(31) \ (48) \ (27,49) \ (28,29,30) \ \ \ \ \mathtt{ffffff}  \ \ \ \
                (31) \ (48) \ (28,30) \ (27,29,49) \ \ \ \ \mathtt{4d303f}$ \newline
\hspace*{15mm} $(31) \ (48) \ (28,49) \ (27,29,30) \ \ \ \ \mathtt{4d303f}  \ \ \ \
                (31) \ (49) \ (27,30) \ (28,29,48) \ \ \ \ \mathtt{4d303f}$ \newline
\hspace*{15mm} $(31) \ (49) \ (27,48) \ (28,29,30) \ \ \ \ \mathtt{ffffff}  \ \ \ \
                (31) \ (49) \ (28,48) \ (27,29,30) \ \ \ \ \mathtt{4d303f}$ \newline
\hspace*{15mm} $(31) \ (49) \ (29,30) \ (27,28,48) \ \ \ \ \mathtt{b2cfc0}  \ \ \ \
                (31) \ (27,30) \ (28,29) \ (48,49) \ \ \ \ \mathtt{340fd3}$ \newline
\hspace*{15mm} $(48) \ (49) \ (27,31) \ (28,29,30) \ \ \ \ \mathtt{ffffff}  \ \ \ \
                (48) \ (49) \ (28,29) \ (27,30,31) \ \ \ \ \mathtt{b2cfc0}$ \newline
\hspace*{15mm} $(48) \ (49) \ (28,31) \ (27,29,30) \ \ \ \ \mathtt{4d303f}  \ \ \ \
                (48) \ (27,30) \ (28,29) \ (31,49) \ \ \ \ \mathtt{340fd3}$ \newline
\hspace*{15mm} $(49) \ (27,30) \ (28,29) \ (31,48) \ \ \ \ \mathtt{340fd3}$

\begin{center}
\large{Table IV}\footnotesize{ Complete GPS$_{x,1}$ DPS set (up to a maximum of 5 PSs)}.
\end{center}

\newpage
{\bf 6. The Hypergraph Links}

Ordinary graphs are made of \textbf{nodes} and \textbf{links} joining two nodes, in \textbf{hypergraphs} links can have 2 or more nodes. As we shall see below this concept is useful in describing the $3$-dimensional structure of molecular DPSs. Before going into concrete examples we need the following definition:

\begin{itemize}
 \item for $c_1 \neq  c_2$
 \item let $\{\text{DPS}\}_{\text{GPS}_{c_1,n_1}}$ be the set 4-DPS associated to GPS$_{c_1,n_1}$ and
 \item let $\{\text{DPS}\}_{\text{GPS}_{c_2,n_2}}$ be the set 4-DPS associated to GPS$_{c_2,n_2}$
\end{itemize}

\textbf{Definition 2} : $\text{DPSa} \in \{\text{DPS}\}_{\text{GPS}_{c_1,n_1}}$ and
                        $\text{DPSb} \in \{\text{DPS}\}_{\text{GPS}_{c_2,n_2}}$ are said to be $c2$-\textbf{compatible} if DPSa$_{c_2}$ and DPSb$_{c_2}$ reduced to the set of common atom numbers are the same.

This concept is crucial for building molecular DPSs, because they are made of the three sets  $\{\mathrm{DPS}_x,  \mathrm{DPS}_y,\mathrm{DPS}_z\}$, an these may not be independent because of shared atom numbers in GPSs from different euclidean dimensions: two DPSs from two atom sharing GPSs in different euclidean dimensions must be compatible if they are to participate in the building of a molecular DPS, otherwise that would generate inconsistent $3D$ structures.

As an example GPS$_{x,1}$ has the following atom numbers in common with other GPSs

\begin{itemize}
 \item 31 and 48 with GPS$_{y,15}$,
 \item 27 and 29 with GPS$_{y,21}$ and
 \item 28, 29 and 31 with GPS$_{z,8}$.
\end{itemize}

this gives the following reduced dominance patterns for the 4-DPS 

\begin{enumerate}
 \item $R_{x,1/y,15}= (31,48)     \text{ on GPS}_{y,15}$
 \item $R_{x,1/y,21}= (29)(27)    \text{ on GPS}_{y,21}$
 \item $Ra_{x,1/z,8}= (28,31)(29) \text{ and } Rb_{x,1/z,8}= (28,29,31) \text{ on GPS}_{z,8}$
\end{enumerate}

where the code $R_{c_1,n_1/c_2,n_2}$ means that the pattern is extracted from some 4-DPS$_{c_2}$ in GPS$_{c_1,n_1}$ and is used to select DPSs in  GPS$_{c_2,n_2}$ that have it as a subpattern.

GPS$_{y,15}$ has only 1 simplex with a set 3 projections (table V), we can also see that any pattern can match  $\{(9,31,41,48)\}_y$ the one and only 4-DPS$_y$ in this GPS.
We can also see from table V that every DPS in GPS$_{x,1}$ can be matched by one of the three patterns that can be extracted from the 4-DPS$_x$s in GPS$_{y,15}$. Thus every DPS in GPS$_{x,1}$ is compatible with at leat one DPS in GPS$_{y,15}$ and reciprocally, in this case the notion of \textbf{compatibility} extends to the whole sets.

\vspace*{4mm}
\hspace*{28mm}  $\{\{(48) \ (31) \ (9) \ (41)\}_x,\{(9,31,41,48)\}_y,\{(9) \ (31) \ (41) \ (48)\}_z\}$ \newline
\hspace*{28mm}  $\{\{(31) \ (48) \ (9) \ (41)\}_x,\{(9,31,41,48)\}_y,\{(9) \ (31) \ (41) \ (48)\}_z\}$ \newline
\hspace*{28mm}  $\{\{(31,48) \ (9) \ (41)\}_x, \ \ \{(9,31,41,48)\}_y,\{(9) \ (31) \ (41) \ (48)\}_z\}$

\begin{center}
\large{Table V} \footnotesize{ Complete DPS projections set for PS$_{y,15}$ DPS}.
\end{center}

\newpage
\hspace*{11mm} $(31) \ (27,48,49) \ (28,29,30)  \ \ \ \ \ \ \mathtt{08}  \ \ \ \ \ \ \ \ \ \
                (29) \ (4,53) \ (7,26,27,52,54)     \ \ \ \ \mathtt{c0}$ \newline
\hspace*{11mm} $(27,31) \ (48,49) \ (28,29,30)  \ \ \ \ \ \ \mathtt{40}  \ \ \ \ \ \ \ \ \ \
                (4,26) \ (7,27) \ (29,52,53,54)     \ \ \ \ \mathtt{04}$ \newline
\hspace*{11mm} $(27,48) \ (31,49) \ (28,29,30)  \ \ \ \ \ \ \mathtt{40}  \ \ \ \ \ \ \ \ \ \
                (4,26) \ (7,27,54) \ (29,52,53)     \ \ \ \ \mathtt{04}$ \newline
\hspace*{11mm} $(27,49) \ (31,48) \ (28,29,30)  \ \ \ \ \ \ \mathtt{40}  \ \ \ \ \ \ \ \ \ \
                (7,27) \ (4,26,53) \ (29,52,54)     \ \ \ \ \mathtt{08}$     \newline
\hspace*{11mm} $(31) \ (48) \ (27,49) \ (28,29,30)  \ \ \ \ \mathtt{0820aa}  \ \ \ \
                (7,27) \ (4,52,53) \ (26,29,54)     \ \ \ \ \mathtt{08}$     \newline
\hspace*{11mm} $(31) \ (49) \ (27,48) \ (28,29,30)  \ \ \ \ \mathtt{0820aa}  \ \ \ \
                (7,52) \ (4,26,27) \ (29,53,54)     \ \ \ \ \mathtt{48}$     \newline
\hspace*{11mm} $(48) \ (49) \ (27,31) \ (28,29,30)  \ \ \ \ \mathtt{0820aa}  \ \ \ \
                (7,52) \ (4,27,53) \ (26,29,54)     \ \ \ \ \mathtt{40}$     \newline
\hspace*{81mm} $(26,29) \ (4,52,53) \ (7,27,54)     \ \ \ \ \mathtt{40}$     \newline
\hspace*{81mm} $(29,52) \ (4,26,53) \ (7,27,54)     \ \ \ \ \mathtt{40}$     \newline
\hspace*{81mm} $(29,52) \ (4,27,53) \ (7,26,54)     \ \ \ \ \mathtt{40}$     \newline
\hspace*{81mm} $(29,54) \ (4,26,27) \ (7,52,53)     \ \ \ \ \mathtt{48}$     \newline
\hspace*{81mm} $(26) \ (4,7) \ (27,53) \ (29,52,54)     \ \ \mathtt{002008}$ \newline
\hspace*{81mm} $(26) \ (4,54) \ (27,53) \ (7,29,52)     \ \ \mathtt{000008}$ \newline
\hspace*{81mm} $(26) \ (7,52) \ (27,53) \ (4,29,54)     \ \ \mathtt{002000}$ \newline
\hspace*{81mm} $(26) \ (27,53) \ (52,54) \ (4,7,29)     \ \ \mathtt{000080}$ \newline
\hspace*{81mm} $(4,7) \ (26,29) \ (27,53) \ (52,54)     \ \ \mathtt{000020}$ \newline
\hspace*{81mm} $(4,52) \ (7,54) \ (26,29) \ (27,53)     \ \ \mathtt{820000}$ \newline
\hspace*{81mm} $(4,54) \ (7,52) \ (26,29) \ (27,53)     \ \ \mathtt{820000}$ \newline  \newline
\hspace*{36mm} \large{\textbf{a)}}
\hspace*{72mm} \large{\textbf{b)}}            \newline

\begin{center}
\large{Table VI} 
\footnotesize{ a) DPSs in a) GPS$_{x,1}$ and b)PS$_{y,21}$ arising from projections matching $R_{x,1/y,21}$ and $Ra_{y,21/x,1}$.}
\end{center}

\vspace*{4mm}

\hspace*{2mm}   $(30) \ (27,28,31) \ (29,48,49) \hspace*{6mm}    \mathtt{04} \hspace*{14mm}
                  (14,17,29) \ (15,32,33) \ (12,28,31,34)            \ \ \ \ \mathtt{10}$     \newline
\hspace*{2mm}   $(48) \ \ (27,29,30) \ (28,31,49)      \ \ \ \ \ \mathtt{08} \hspace*{14mm}
                  (12) \ (14) \ (28,31,32) \ (15,17,29,33,34)        \ \ \ \ \mathtt{082000}$ \newline
\hspace*{2mm}   $(28,31) \ (48,49) \ \ (27,29,30)    \ \ \ \ \ \ \mathtt{10} \hspace*{14mm}
                  (12) \ (17,33) \ (28,31) \ (14,15,29,32,34)        \ \ \ \ \mathtt{082000}$ \newline
\hspace*{2mm}   $(29,48) \ (30,49) \ \ (27,28,31)    \ \ \ \ \ \ \mathtt{40} \hspace*{14mm}
                  (12) \ (28,31) \ (33,34) \ ( 14,15,17,29,32)       \ \ \ \ \mathtt{082000}$ \newline
\hspace*{2mm}   $(30) \ (49) \ (29,48) \ (27,28,31)  \ \ \ \ \mathtt{0820aa} \hspace*{6mm}
                  (12) \ (14,15,17) \ (28,31,32) \ (29,33,34)        \ \ \ \ \mathtt{410000}$ \newline
\hspace*{2mm}   $(48) \ (49) \ (28,31) \ (27,29,30)  \ \ \ \ \mathtt{041055} \hspace*{6mm}
                  (12) \ (14,17,29) \ (15,32,33) \ (28,31,34)        \ \ \ \ \mathtt{082000}$ \newline
\hspace*{115mm} $.$ \newline
\hspace*{115mm} $.$ \newline
\hspace*{115mm} $.$ \newline
\hspace*{73mm}  $(28,31) \ (33,34) \ (12,29,32) \ (14,15,17) \ \ \ \ \mathtt{2c1000}$ \newline
\hspace*{73mm}  $(29,32) \ (33,34) \ (12,28,31) \ (14,15,17) \ \ \ \ \mathtt{00a0c0}$ \newline
\hspace*{73mm}  $(29,33) \ (32,34) \ (12,28,31) \ (14,15,17) \ \ \ \ \mathtt{000800}$ \newline  \newline
\hspace*{36mm} \large{\textbf{a)}}
\hspace*{72mm} \large{\textbf{b)}} \newline

\begin{center}
\large{Table VII}
\footnotesize{ DPSs in a) GPS$_{x,1}$ and b) GPS$_{z,8}$ arising from projections matching $Ra_{x,1/z,8}$ and $Ra_{z,8/x,1}$. The complete set b) contains 100 DPSs, only the first and the last few are shown.}
\end{center}

\newpage
Likewise a set of constraints arise from GPS$_{y,15}$, GPS$_{y,21}$ and GPS$_{z,8}$ that must be fulfilled by GPS$_{x,1}$

$\mathbf{GPS}_{y,15}$ : $Pa_{y,15/x,1}= (31)( 48) \text{ , } Pb_{y,15/x,1}= (48)( 31) \text{ and }                                                                          Pc_{y,15/x,1}= (31,48)$
 \hspace*{9mm}  (6a) \newline
$\mathbf{GPS}_{y,21}$ : $Pa_{y,21/x,1}= (27)(29) $
\hspace*{93.4mm}  (6b) \newline
$\mathbf{GPS}_{z,8}$ \ : $Pa_{z,8/x,1}=(29)(28,31) \text{ and } Pa_{z,8/x,1}=(28,29,31)$
\hspace*{40mm}  (6c)

By sorting reciprocally compatible DPSs with these patterns we have been able to construct the links between DPS sets:

\begin{enumerate}
 \item GPS$_{x,1}$ and GPS$_{y,15}$, tables IV and V respectively.
 \item GPS$_{x,1}$ and GPS$_{y,21}$, tables VIa and VIb respectively.
 \item GPS$_{x,1}$ and GPS$_{z,8}$, tables VIIa and VIIb respectively.
\end{enumerate}

which completes our example.

\vspace*{4mm}
{\bf 7. Conclusion}

This paper shows that it is algorithmically feasible to enumerate the sets of cells in $CS$ that are dynamically accessible for a thermalised molecule, and store the result in a compact format. This is possible because the generalized partition sequences allow a second factorization of DPSs, thus enabling a high level of data compaction. Further compaction is still possible and will be explored in future works.

It should not be forgotten that the present results have necessitated writing a lengthy computer program with 20000 lines of code and, more important, 350000 lines of miscellaneous data distributed among a set of 325 arrays.

\newpage
{\bf Appendix}

List of abbreviations used in this work

$\begin{array}{ll}
       CS         & \text{conformational space}                     \\
 \text{MDS}       & \text{molecular dynamics simulation}            \\
 \text{DPS}       & \text{dominance partition sequence}             \\
 \text{DPS}_c     & \text{dominance partition sequence for the } c \text{ coordinate axis}      \\
 \text{GDPS}      & \text{generalized dominance partition sequence} \\
 \text{GPS}       & \text{generalized permutation sequence}         \\
 \text{GPS}_{c,n} & \text{n}^{\text{th}}
                    \text{ generalized permutation sequence for the } c \text{ coordinate axis} \\
 \text{PS}        & \text{permutation sequence}                     \\
\end{array}$

\newpage
{\bf References}

\begin{itemize}

\item[[1]]  J. Gabarro-Arpa,
            A central partition of molecular conformational space. I. Basic structures,
            Comp. Biol. and Chem.
            27 (2003) 153-159.

\item[[2]]  J. Gabarro-Arpa,
            A central partition of molecular conformational space. II. Embedding 3D-structures,
            in
           {\it Proceedings of the 26th Annual International Conference of the IEEE EMBS},
           (San Francisco 2004), pp. 3007-3010.

\item[[3]]  J. Gabarro-Arpa,
            Combinatorial determination of the volume spanned by a molecular system in conformational space,
            Lecture Series on Computer and Computational Sciences
            4 (2005) 1778-1781.

\item[[4]]  J. Gabarro-Arpa,
            A central partition of molecular conformational space. III.
            Combinatorial determination of the volume spanned by a molecular system in conformational space,
            Journal of Mathematical Chemistry
            42 (2006) 691-706.

\item[[5]]  J. Gabarro-Arpa,
            Heuristic decomposition of cones in molecular conformational space.
            physics.comp-ph:0710.2529v1 (2007)

\item[[6]]  J. Gabarro-Arpa,
            A central partition of molecular conformational space. IV.
            Extracting information from the graph of cells,
            J. Math. Chem. 44 (2008) 872–883.

\item[[7]]  P.G. Mezey
            {\it Potential Energy Hypersurfaces},
            (Elsevier, Amsterdam, 1987).

\item[[8]]  D.J. Wales,
            {\it Energy Landscapes},
            (Cambridge University Press, Cambridge, 2003).

\item[[9]]  K. Henzler-Wildman, D. Kern,
            Dynamic personalities of proteins,
            Nature
            450 (2007) 964-972.

\item[[10]] J. Gabarro-Arpa,
            Embbeding protein 3D-structures in a cubic lattice. I. The basic algorithms.
            physics: 1004.2022v1 (2010)

\item[[11]] S. Fomin and N. Reading, Root systems and generalized associahedra,
            math.CO/0505518 (2005).

\item[[12]] C.R. Shalizi and C. Moore,
            What is a macrostate? Subjective observations and objective dynamics,
            cond-mat/0303625 (2003).

\item[[13]] B. Liskov and J. Wing,
            A Behavioural Notion of Subtyping,
            ACM Transactions on Programming Languages and Systems 16 (1994) 1811-1841.

\item[[14]] R. Br\"uggemann, L. Carlsen,
            {\it Partial Order in Environmental Sciences and Chemistry},
           (Springer, Berlin, 2006).

\item[[15]] G. Restrepo, R. Br\"uggemann,
            K. Voigt, Croat. Chem. Acta. 80 (2007) 261.

\item[[16]] G. Restrepo, R. Br\"uggemann,
            J. Math. Chem. 44 (2008) 577–602.

\item[[17]] J.R. Dias,
            J. Math. Chem. 4 (1990) 17.

\item[[18]] E.E. Daza, A. Bernal,
            J. Math. Chem. 38 (2005) 247.

\item[[19]] A. Bernal,
            {\it Ordenamientos moleculares basados en la energía}
            (BSc Thesis, Universidad Nacional de Colombia, Bogotá, 2004).

\item[[20]] Papers in MATCH Commun. Math. Comput. Chem. 42,7 (2000); 54, 489 (2005).

\item[[21]] D.J. Klein,
            J. Math. Chem. 18, 321 (1995).

\item[[22]] D.J. Klein, D. Babi
            J. Chem. Inf. Comput. Sci. 37, 656 (1997).

\item[[23]] A.D. MacKerell Jr., et al.,
            All-Atom empirical potential for molecular modeling and dynamics studies of proteins,
            J. Phys. Chem. B
            102 (1998) 3586-3616.

\item[[24]] W. Wang, O. Donini, C.M. Reyes, P.A. Kollman,
            Biomolecular simulations: recent developments in force fields, simulations of enzyme catalysis,
            protein-ligand, protein-protein, and protein-nucleic acid noncovalent interactions,
            Annu. Rev. Biophys. Biomol. Struct.
            30 (2001) 211-243.

\item[[25]] M. Marquart, J. Walter, J. Deisenhofer, W. Bode, R. Huber,
            The geometry of the reactive site and of the peptide groups in trypsin,
            trypsinogen and its complexes with inhibitors,
            Acta Crystallogr. Sect. B
            39 (1983) 480-490.

\item[[26]] H.S.M. Coxeter,
            {\it Regular polytopes},
            (Dover Publications Inc., New York, 1973).

\item[[27]] A. Bjorner, M. las Vergnas, B. Sturmfels, N. White,
            {\it Oriented Matroids},
            (Cambridge University Press, Cambridge, UK, sect. 2, 1993).

\item[[28]] J. Gabarro-Arpa, R. Revilla,
            Clustering of a molecular dynamics trajectory with a Hamming distance,
            Comp. and Chem.
            24 (2000) 693-698.

\item[[28]] K.H. Rosen Ed. in Chief
            {\it Handbook of Discrete and Combinatorial Mathematics}
            (CRC Press, Boca Raton, USA, chap. 8 sect 12, 2000).

\end{itemize}

\end{document}